\documentclass[pre,aps,twocolumn]{revtex4}

\usepackage{graphicx}
\usepackage{amsmath}
\usepackage{amssymb}
\usepackage{ifthen}
\usepackage{booktabs}

\usepackage{graphicx}
\usepackage{dcolumn}
\usepackage{bm}
\usepackage{longtable}

\newcommand{\bb}{\begin{equation}}
\newcommand{\ee}{\end{equation}}
\newcommand{\ba}{\begin{eqnarray*}}
\newcommand{\ea}{\end{eqnarray*}}
\newcommand{\rhor}{\rho({\bf r})}
\newcommand{\dd}{{\rm d}}
\newcommand{\rr}{{\mathbf r}}
\newcommand{\dr}{{\rm d}{\bf r}}

\begin{document}

\title{Critical Point Wedge Filling}

\author{Alexandr \surname{Malijevsk\'y}}
\affiliation{
{Department of Physical Chemistry, Institute of Chemical Technology, Prague, 166 28 Praha 6, Czech Republic}}
\author{Andrew O. \surname{Parry}}
\affiliation{Department of Mathematics, Imperial College London, London SW7 2B7, UK}

\begin{abstract}
We present results of a microscopic density functional theory study of wedge filling transitions, at a right-angle wedge, in the presence of dispersion-like wall-fluid forces. Far from the corner the walls of the wedge show a
first-order wetting transition at a temperature $T_w$ which is progressively closer to the bulk critical temperature $T_c$ as the strength of the wall forces is reduced. In addition, the meniscus formed near the corner
undergoes a filling transition at a temperature $T_f<T_w$, the value of which is found to be in excellent agreement with macroscopic predictions. We show that the filling transition is {\it first-order} if it occurs far from
the critical point but is {\it continuous} if $T_f$ is close to $T_c$ even though the walls still show first-order wetting behaviour.  For this continuous transition the distance of the meniscus from the apex grows as
$\ell_w\approx (T_f-T)^{-\beta_w}$ with critical exponent  $\beta_w\approx 0.46 \pm 0.05$ in good agreement with the phenomenological effective Hamiltonian prediction. Our results suggest that critical filling transitions,
with accompanying large scale universal interfacial fluctuation effects, are more generic than thought previously, and are experimentally accessible.

\end{abstract}

\maketitle

There is now direct experimental evidence for the thermal excitation of the gravity stabilised capillary-wave-like fluctuations at the interface between coexisting fluid phases \cite{Dirk}. Over the last few decades theory has
predicted that such fluctuation effects are particularly important at certain types of interfacial phase transition such as critical wetting \cite{dietrich, sullivan, schick, bonn, saam}. Wetting refers to the unbinding of a
fluid interface from a solid substrate (or another fluid interface) on approaching a temperature $T_w$, at which the contact angle $\theta$ vanishes. The order of these transitions is determined by the subtle interplay between
wall-fluid and fluid-fluid intermolecular forces and also interfacial fluctuations. The original macroscopic argument for wetting transitions had predicted that the transition would be first-order, and should necessarily occur
on approaching the bulk critical temperature $T_c$ \cite{Cahn}. Model calculations soon revealed that the location and order of the transition are more general than this. In particular, Nakanishi and Fisher showed that, for
systems with short-ranged forces, the transition should change from first-order to continuous, if the surface forces are weakened and $T_w$ approaches $T_c$ \cite{nakanishi}. While this has been fully tested in Ising model
studies \cite{Binderwet}, this scenario is altered by the presence of long-ranged, dispersion-like intermolecular forces. In order to see continuous (now referred to as ``critical") wetting transitions one requires a fine
tuning of the range and strengths of the solid-fluid and fluid-fluid forces \cite{dietrich, lipowsky}.  Consequently while there are many examples of first-order wetting, there are no unambiguous experimental examples of
critical wetting for solid-fluid interfaces,  although the transition has been seen in a few binary mixtures \cite{bonn}.

One way around this, which would allow one to see the strong influence of interfacial fluctuations on a continuous phase transition, is to consider fluid adsorption in a linear wedge for which there is an analogous transition
referred to as filling \cite{shuttle, concus, pomeau, hauge, rejmer}. This transition is far more common in nature than the wetting transition and was first studied experimentally almost 40 years ago \cite{finn} although the
order of the transition was not considered. Recent phenomenological effective Hamiltonian models have predicted that fluctuation effects are enhanced compared to wetting and also that the requirements that the transition can
be continuous are more relaxed \cite{wood1, wood2, bernardino}. While this has been studied extensively in the Ising model \cite{binder03,binder05,parry05, abraham02,abraham03}, the more realistic case of long-ranged forces
has not been studied in detail. In this paper we present the results of a study of filling in the presence of dispersion forces, based on a microscopic classical density functional theory (DFT). The latter has been
instrumental in developing our understanding of inhomogeneous fluids but is most usually applied to systems in which the equilibrium density depends on only one co-ordinate \cite{evans86, evans90, stewart, nold}. Here we use a
2D DFT to study filling transitions and compare with the predictions of thermodynamic arguments and effective Hamiltonian theory. We find that close to $T_c$ the filling transition is continuous even though the walls of the
wedge themselves still exhibit first-order wetting.  This allows us to check interfacial Hamiltonian predictions for the critical behaviour and is strongly encouraging that continuous filling transitions may be found in the
laboratory similar to experiments on complete wedge filling \cite{mistura}.

Consider a wedge geometry formed by two identical infinite planar walls that meet at an opening angle $ 2\psi$ in contact with a bulk vapour at chemical potential $\mu$, tuned to saturation $\mu=\mu_{\rm sat}^-$ at a
temperature $T<T_c$. The wedge may be thought as being a missing link between a planar wall ($\psi=\pi/2$) and a  capillary-slit ($\psi=0$) and shows a phase transition, filling, which is distinct from wetting and capillary
condensation. Far from the apex the thickness of the liquid wetting layer, $\ell_\pi$, is the same as for a planar wall. However near the apex, the thickness of the meniscus can be much greater. Macroscopic arguments
 dictate that the wedge is completely filled above a filling transition temperature $T_f$ which occurs when the contact angle of a liquid drop satisfies \cite{shuttle, concus, pomeau, hauge}
\begin{equation}
 \theta(T_f)=\frac{\pi}{2}-\psi\,.
 \label{thermo}
\end{equation}
 The {\it{wedge filling transition}} corresponds to the change from microscopic to macroscopic adsorption, as $T\to T_f$, and may be
first-order or continuous (critical filling) corresponding to the discontinuous or continuous divergence of the adsorption.  Because (\ref{thermo}) is an exact requirement, the filling transition is ubiquitous in nature for
all fluids that have form drops with a finite contact angle.

Within classical DFT the equilibrium density profile is found by minimizing the grand potential functional $ \Omega[\rho]=F[\rho]+\int \dr\rhor[V(\rr)-\mu]$, where $V(\rr)$ is the external potential \cite{evans_79}. We
consider a right angle wedge ($\psi=\pi/4$) so that the potential $V({\bf{r}})=V(x,z)$ is a function of Cartesians $x,z>0$ and is translationally invariant along the wedge. Here $F[\rho]$ is the intrinsic free energy
functional of the fluid one-body density, $\rho(\bf{r})$, which can be split into ideal and excess parts. Modern DFT often divides the latter into a hard-sphere part $F_{\rm hs}[\rho]$ and an attractive contribution
$F_{a}[\rho]=\frac{1}{2}\int\int \dr_1 \dr_2\rho(\rr_1)\rho(\rr_2)u_{a}(r_{12})$ where  $u_{a}(r)$ is the attractive part of the fluid-fluid potential. We take this to be a Lennard-Jones (LJ) potential $u_{a}(r)= -4\varepsilon
(\sigma/r)^6H(r-\sigma)$ which is truncated at $r_c=2.5\,\sigma$, where $\sigma$ is the hard-sphere diameter and $H(x)$ is the Heaviside function. For $F_{\rm hs}[\rho]$ we use Rosenfeld's fundamental theory which accurately
models packing effects if the density is high close to the walls \cite{ros, roth_fmt}. The external potential arises from a uniform distribution of wall atoms, with density $\rho_w$, which for $r>\sigma$  interact with the
fluid atoms via the LJ potential $\phi_w(r)=-4\varepsilon_w\left(\frac{\sigma}{r}\right)^{6}$, leading to
 \begin{equation}
V(x,z)=\alpha_w\left[\frac{1}{z^3}+\frac{2z^4+x^2z^2+2x^4}{2x^3z^3\sqrt{x^2+z^2}} +\frac{1}{x^3}\right]\,,
 \end{equation}
where $\alpha_w=-\frac{1}{3}\pi\varepsilon_w\rho_w\sigma^6$. There is a hard-wall repulsion if $x,z<\sigma$. Infinitely far from the wedge apex,  the
potential close to either surface recovers that of a planar wall e.g. $V(\infty,z)=2\alpha_w/z^3$. The functional $\Omega[\rho]$ is minimized numerically on an
$L\times L$ grid where the lateral dimension of our box is $L=50\sigma$ and the grid has discretization size $0.05\,\sigma$. To mimic the bulk boundary
conditions we impose $\rho(L,z)=\rho_\pi(z)$ and $\rho(x,L)=\rho_\pi(x)$ where $\rho_\pi(z)$ is the equilibrium profile for a planar wall-fluid interface with
$\rho_\pi(L)$ fixed to the bulk gas density $\rho_g$. In our model DFT $k_BT_c/\varepsilon=1.414$ and temperature is expressed either in fractions of $T_c$ or in dimensionless units $T^*=k_BT/\varepsilon$.

\begin{figure}
\includegraphics[width=0.45\textwidth]{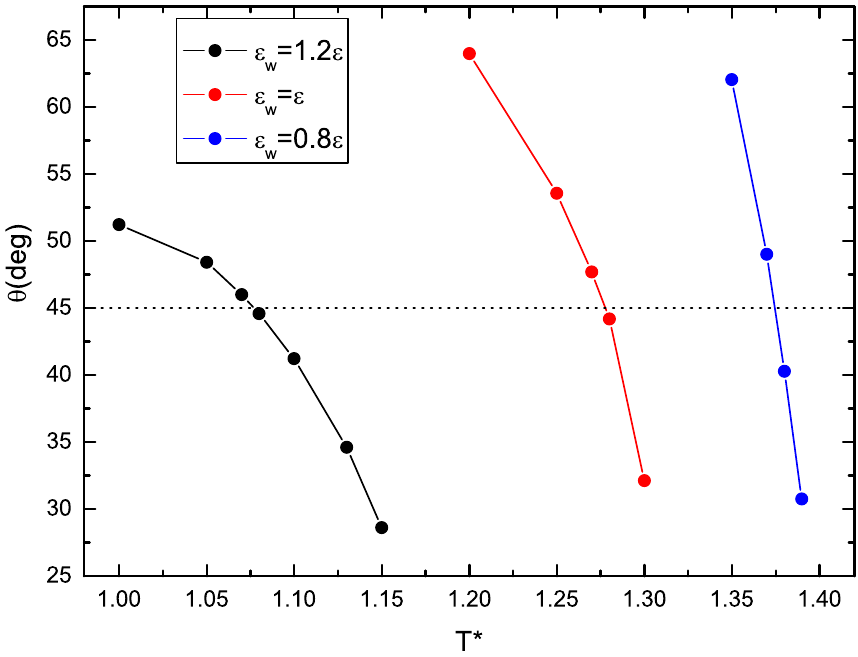}
\caption{Variation of the contact angle with $T$ for different wall strengths. The intersection with the dashed line at $\theta=45^\circ$ is the thermodynamic prediction for $T_f$. }\label{cont_angle}
\end{figure}

\begin{figure}
 \includegraphics[width=0.45\textwidth]{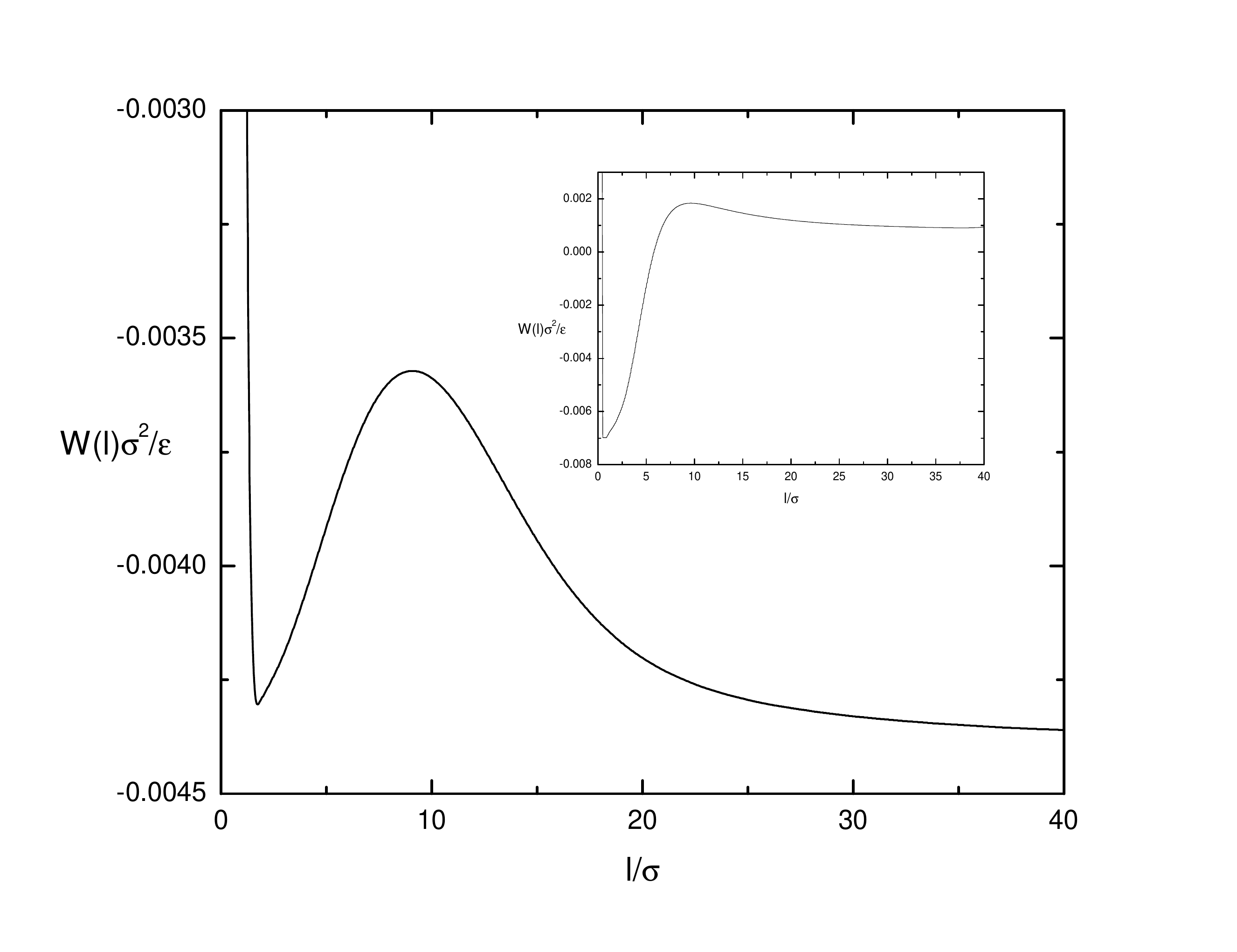}
\caption{Binding potential function $W(\ell)$ for $\varepsilon_w=0.8\,\varepsilon$ close to a first-order wetting transition at $T_w^*=1.4$  showing an activation barrier at $\ell_B\approx 10\,\sigma$. In the inset we show the
binding potential at a lower temperature close to $T_f^*\approx 1.38$ for which the activation barrier is still present.  In both cases the results correspond to a bulk coexistence.}\label{bind08}
\end{figure}

We have considered a variety of wall strengths and present results for $\varepsilon_w=1.2\,\varepsilon$, $\varepsilon_w=\varepsilon$ and $\varepsilon_w=0.8\,\varepsilon$. For each, we first considered the planar wall with
potential $V_\pi (z)=2\alpha_w /z^3 $ and determined the density profile $\rho_\pi(z)$ and surface tensions $\gamma_{wg}$, $\gamma_{wl}$ and $\gamma$ of the wall-gas, wall-liquid and liquid-gas interfaces, respectively. From
Young's equation $\cos\theta=(\gamma_{wg}-\gamma_{wl})/\gamma$ we determined $\theta(T)$ for each of these systems (see Fig.\,\ref{cont_angle}). Each system exhibits a wetting transition, with $T_w$
determined from the crossing of $\gamma_{wg}$ and $\gamma_{wl}+\gamma$. These occur at $T_w=0.83\,T_c$, $T_w=0.93 \,T_c$ and $T_w=0.99\,T_c$ as $\varepsilon_w$ is reduced. The wetting
transitions are all first-order; that is the thickness of the liquid layer $\ell_\pi$ jumps from a microscopic to macroscopic value at $T_w$. This is to be expected since the wall-fluid potential is long-ranged but the truncated LJ fluid-fluid interaction is effectively short-ranged \cite{dietrich}. This prohibits critical wetting, which is important for our study. Also, as expected, the strength of the first-order transition decreases as $T_w$ approaches $T_c$. This is apparent when one determines the interfacial binding interfacial binding potential $W(\ell)$
corresponding to the excess grand potential of a wetting film constrained to be of thickness $\ell$. The global minimum of this determines the equilibrium film thickness $\ell_\pi$. This is shown in Fig.\,\ref{bind08} for the
case $\varepsilon_w=0.8\,\varepsilon$ close to the wetting temperature and shows an activation barrier, characteristic of  first-order wetting, at $\ell_B\approx 10\,\sigma$. For comparison the barrier for the binding
potential for $\varepsilon_w=\varepsilon$ is an order of magnitude larger and located at $\ell_B\approx 4\sigma$.

\begin{figure}
\includegraphics[width=0.45\textwidth]{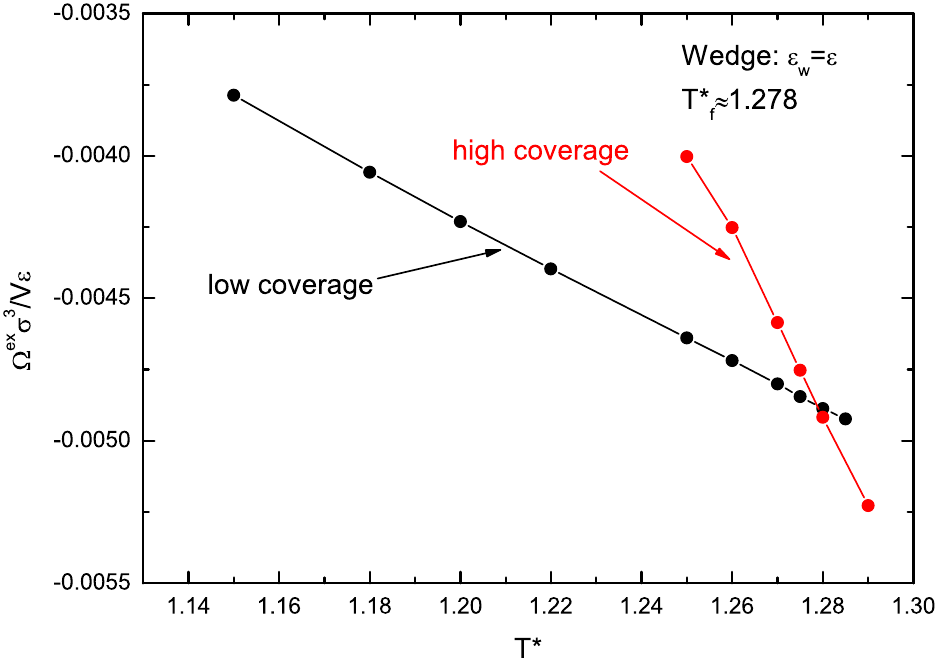}
\caption{Location of a first-order filling transition for $\varepsilon_w=\varepsilon$. Here
$V$ is the available volume which is the length of the wedge multiplied by $(L-\sigma)^2$. }\label{w_1}
\end{figure}

\begin{figure}
\includegraphics[width=0.24\textwidth]{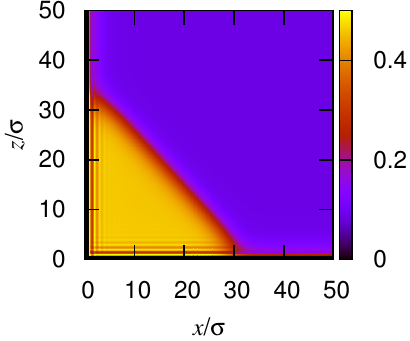} \includegraphics[width=0.24\textwidth]{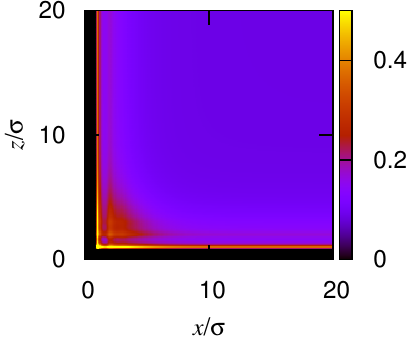}
\caption{Coexisting macroscopic (left) and microscopic (right) density profiles at a first-order filling transition for $\varepsilon_w=\varepsilon$.}\label{pw1_1.28}
\end{figure}

According to the thermodynamic prediction (\ref{thermo}), the location of the filling transitions can be determined from the intersection of the contact angle curves with $\psi=\pi/4$ and gives $T_f=0.76\,T_c$, $T_f=0.90\,T_c$
and $T_f=0.97\,T_c$ as $\varepsilon_w$ decreases in strength. To check this we set $\mu=\mu_{\rm sat}^-$ and minimize $\Omega[\rho]$ to a global or local minimum $\Omega$, starting from different high density and low density
configurations. For first-order filling these will converge to different equilibrium profiles, corresponding to microscopic and macroscopic adsorptions, which coexist at $T_f$. This is what is found for the two strongest walls
as illustrated in Fig.\ref{w_1} where we plot the excess grand-potential $\Omega^{\rm ex}=\Omega+pV$ per unit volume as a function of $T$. The values for $T_f$ obtained are in near exact agreement with the thermodynamic
predictions and differ from them only due to the limitations of numerical discretization and finite-size. In Fig.\,\ref{pw1_1.28} we show the coexisting density profiles $\rho(x,z)$, corresponding to macroscopic (left) and
microscopic (right) states, for $\varepsilon_w=\varepsilon$. From these we can determine the thickness $\ell_w$ of the meniscus above the wedge apex defined as the distance from the origin to a point on a diagonal where
$\rho(x,x)=(\rho_l+\rho_g)/2$. The macroscopic meniscus is nearly flat (as it should be since we are at bulk coexistence) and meets each wall at the correct contact angle $\theta\approx\pi/4$. Of course the size of  this
macroscopic state is limited by our numerical grid and scales with the system size $L$. For the microscopic configuration the meniscus thickness $\ell_w$ is larger than the wetting layer thickness $\ell_\pi$ but of the same
order as the distance of the activation barrier $\ell_B\approx 4\sigma$ for the corresponding binding potential for the wetting transition. This is precisely the expectation for first-order filling from effective Hamiltonian
theory \cite{wood2}. Both microscopic and macroscopic profiles show layering behaviour close to the apex.

\begin{figure}
\includegraphics[width=0.45\textwidth]{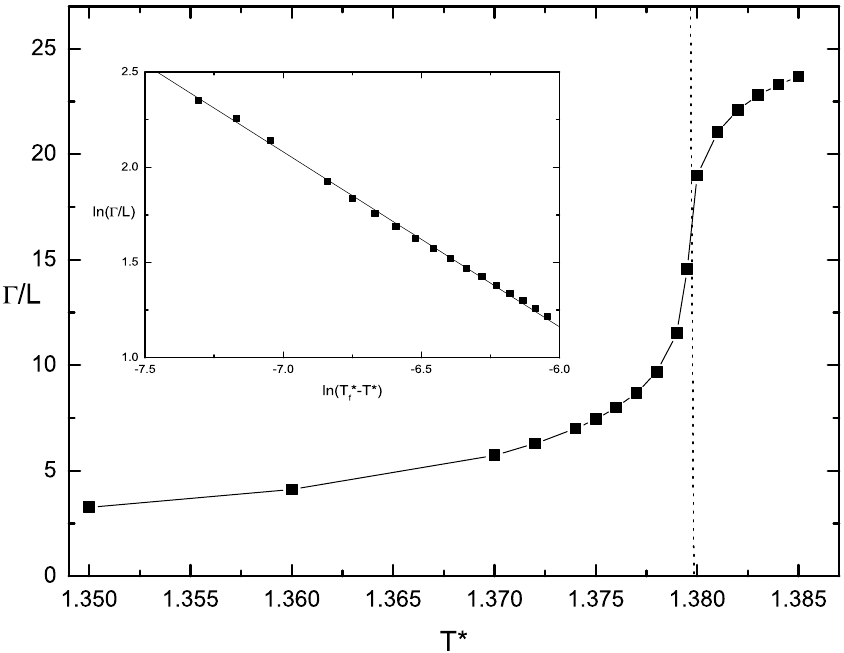}
\caption{Temperature dependence of the adsorption in the wedge with the weakest wall interaction $\varepsilon_w=0.8\,\varepsilon$. The inset shows the log-log
plot of the adsorption vs the scaling field $T_f-T$. The slope of the straight line is $-0.92$. }\label{ads_w08}
\end{figure}

\begin{figure}
\includegraphics[width=0.45\textwidth]{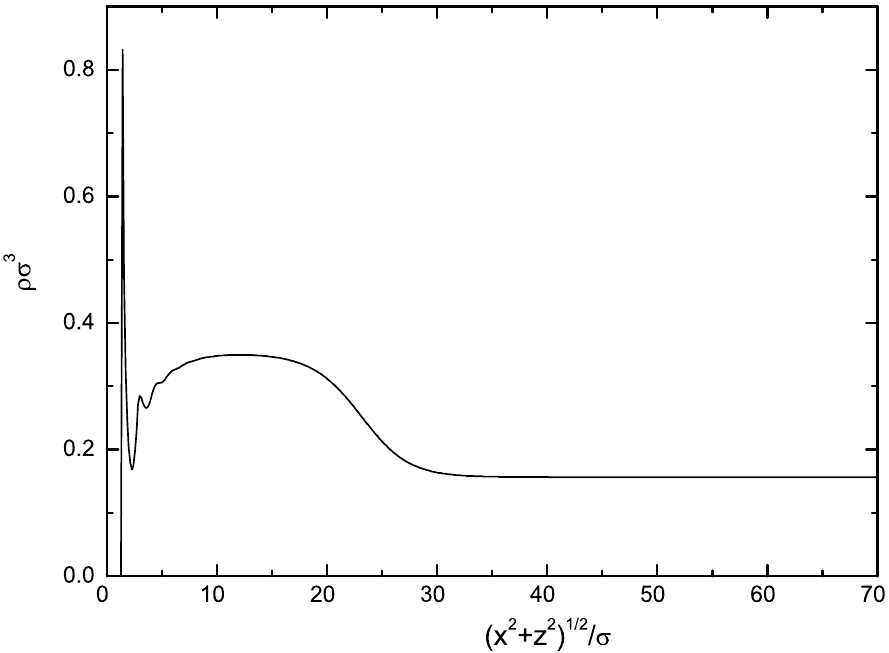}
\caption{Density profile along the diagonal from $(0,0)$ to $(L,L)$ at the filling temperature $T_f$ for $\varepsilon_w=0.8\,\varepsilon$. }
\label{rhoxx}
\end{figure}

For the weakest wall strength $\varepsilon_w=0.8\,\varepsilon$ however both initial coverages converge to a unique phase indicating that the transition, which is of course rounded by the finite-size of our system, is
continuous. A plot of the adsorption $\Gamma=\int\int \dd x \dd z (\rho(x,z)-\rho_g)$ versus $T$ is shown in Fig.\,\ref{ads_w08} and shows a dramatic but continuous increase in the adsorption near the anticipated
$T_f^*\approx1.38$. A cross-section of the density profile along the diagonal, $\rho(x,x)$, for $T\approx T_f$ is also shown (Fig.\,\ref{rhoxx}).
 This indicates that the order of the filling transition is changed near the vicinity of the $T_c$. Support for this comes from the two sources. Firstly near $T_f$ the meniscus height $\ell_w\approx 22\sigma$ is considerably larger than the location of the activation barrier $\ell_B\approx
10\,\sigma$ associated with the wetting binding potential (see inset Fig.\,\ref{bind08}). This is not at all expected for first-order filling \cite{wood2}. Secondly we can compare quantitatively with predictions for critical
filling. If the transition is critical then in an infinite wedge we expect that $\ell_w\sim(T_f-T)^{-\beta_w}$ with $\Gamma\propto \ell_w^ 2$ owing to the triangular shape of the meniscus. Effective Hamiltonian theory predicts
that the critical singularities depend on the power-law describing the dominant wall-fluid or fluid-fluid interaction which we may write more generally as $V(z)\approx 1/z^{p+1}$. The critical behaviour falls into two regimes
with $\beta_w=1/p$ for $p<4$ and $\beta_w=1/4$ for $p>4$ \cite{wood2}. Thus we anticipate $\beta_w=1/2$ in our model since $p=2$. The inset in Fig.\,\ref{ads_w08} shows a log plot of the adsorption for $T<T_f$, in which we use
an {\it{unfitted}}  estimate of the filling temperature $T^*_f=1.38$ obtained from (\ref{thermo}). This gives $\beta_w=0.46 \pm 0.05$ in good agreement with the predicted value.

The presence of a critical (or at least effectively critical) filling transition when $T_f$ is close to $T_c$, when the walls still exhibit first-order wetting, and in the presence of  realistic long-ranged interactions is the
main new result of our paper, and is we believe encouraging for experimental studies. To date there have only been detailed laboratory studies of {\it{complete filling}} in linear wedges corresponding to the approach to
co-existence when the walls are completely wet ($\theta=0$) \cite{mistura}. However the observation of critical filling would be more interesting because fluctuation effects are much stronger. For example, we expect that
beyond mean-field level, capillary-wave fluctuations do not alter the divergence of $\ell_w\approx (T_f-T)^{-1/2}$ but do lead to a universal interfacial roughness (width) $\xi_{\perp}\approx (T_f-T)^{-1/4}$ which is much
bigger than for complete filling and also critical wetting \cite{wood2}. These fluctuations are not captured by the present DFT and consequently the density profiles will be broader than predicted here. However our DFT should
be otherwise extremely accurate regarding the location of the transition, its order and the adsorption. The observed change in order from first-order to continuous filling has only been partially anticipated by previous
effective Hamiltonian theory. This had been predicted on the basis of a simple interfacial model, valid only for {\it{shallow}} wedges, but the proposed mechanism required that both the wall-fluid and fluid-fluid forces to be
of the same range. Then it was noted that even for first order wetting, the filling transition would be continuous if it occurs at a temperature below which the activation barrier forms in the binding potential $W(\ell)$
\cite{wood2}. However in the present DFT study a small activation barrier is still present at $T_f$ (see inset Fig.\,\ref{bind08}), indicating that the prediction of the simple, shallow wedge, effective Hamiltonian theory is
not completely correct. Nevertheless we believe that the substantial reduction in the size of the barrier as $T$ approaches $T_c$ plays a prominent role in the change of order of the filling transition.  Finally it would be
interesting to know if the change in order occurs via a tricritical or critical end point and also what happens for more acute wedges with stronger wall potentials.

In this paper we have presented our results of numerical studies of first-order and critical filling transitions in a rectangular wedge using a non-local density functional theory. This is the first time that filling
transitions have been studied using modern microscopic DFT in the presence of long-ranged wall-fluid interactions, and show that close to the bulk critical temperature the wedge filling transition may be continuous even though
the walls themselves exhibit first-order wetting.

\end{document}